\documentstyle[aas2pp4]{article}
\newcommand{\groj}{GRO~J1744$-$28}
\newcommand{\xte}{{\it RXTE}}

\newcommand{\exosat}{{\it EXOSAT}}

\begin{document}
\title{Post-Burst Quasi-Periodic Oscillations from \groj\ \\ and from the
  Rapid Burster}

\author{Jefferson M.\ Kommers, Derek W.\ Fox, 
  \& Walter H.\ G.\ Lewin}
\affil{Department of Physics and Center for Space Research,\\
  Massachusetts Institute of Technology, Cambridge, MA 02139}

\author{Robert E.\ Rutledge} 
\affil{Max-Planck-Institut f\"{u}r Extraterrestrische Physik, D-85740
  Garching, Germany}

\author{Jan van Paradijs}
\affil{University of Alabama in Huntsville, Huntsville, AL 35812\\
  and University of Amsterdam, Amsterdam, The Netherlands}

\author{Chryssa Kouveliotou}
\affil{Universities Space Research Association, Huntsville, AL 35800}

\slugcomment{arch-ive/9703130}
\begin{abstract}
  The repetitive X-ray bursts from the accretion-powered pulsar \groj\ 
  show similarities to the type II X-ray bursts from the Rapid
  Burster.  Several authors (notably Lewin et al.) have suggested that
  the bursts from \groj\ are type II bursts (which arise from the
  sudden release of gravitational potential energy).  In this paper,
  we present another similarity between these sources.  {\it Rossi
    X-ray Timing Explorer} observations of \groj\ show that at least
  10 out of 94 bursts are followed by quasi-periodic oscillations
  (QPO) with frequencies of $\sim$ 0.4 Hz.  The period of the
  oscillations decreases over their $\sim 30$--80 s lifetime, and
  they occur during a spectrally hard ``shoulder'' (or ``plateau'')
  which follows the burst.  In one case the QPO show a modulation
  envelope which resembles simple beating between two narrow-band
  oscillations at $\sim$ 0.325 and $\sim$ 0.375 Hz.  Using \exosat\ 
  observations, Lubin et al. found QPO with frequencies of 0.039 to
  0.056 Hz following 10 out of 95 type II bursts from the Rapid
  Burster.  As in \groj\, the period of these oscillations decreased
  over their $\sim 100$ s lifetime, and they occurred only during
  spectrally hard ``humps'' in the persistent emission.  Even though
  the QPO frequencies differ by a factor of $\sim$ 10, we believe that
  this is further evidence that a similar accretion disk instability
  is responsible for the type II bursts from these two sources.
\end{abstract}

% I selected keywords from the 1996 ApJ keyword list
\keywords{X-rays: general, bursts --- stars: neutron}

\section{Introduction}
\groj\ is a recently discovered accretion-powered pulsar which shows
repetitive X-ray bursts (\cite{KouveliotouNature96};
\cite{StrickmanApJLett96}; \cite{GilesApJLett96}; and references
therein).  Its unusual bursting behavior has been compared (see below)
to that of the Rapid Burster (MXB 1730$-$335), which was discovered 20
years ago by Lewin et al. (1976).  Analysis of the X-ray bursts from
the Rapid Burster showed that it produced two distinct types of burst:
type I, attributed to thermonuclear flashes on the surface of a
neutron star; and type II, attributed to the release of gravitational
potential energy due to spasmodic accretion.  The mechanism
responsible for the type II bursts has not been fully understood,
although it is almost certainly related to an accretion disk
instability (for reviews see Lewin, van Paradijs, \& Taam 1993,1995).
Rapidly repetitive type II bursts had previously been observed only
from the Rapid Burster, so if the bursts from \groj\ are also of type
II---as convincingly argued by Lewin et al. (1996)---then a comparison
of these sources may constrain theories of the burst mechanism.

The pulse period of \groj\ is 467 ms and the neutron star is in a 11.8
day binary orbit about a low-mass donor star (\cite{FingerNature96}).
During the first 12 hours of bursting observed with BATSE on December
2, 1995, bursts occurred every 3 to 8 minutes, and for one three hour
period the burst intervals clustered around $172 \pm 15$ s
(\cite{KouveliotouNature96}).  Subsequently the burst intervals became
longer and more erratic and the burst rate settled at about 30--40 per
day (corrected for Earth occultation and live time;
\cite{FishmanIauc6290}).  Burst durations were initially $\sim$ 20--30
s, and settled down to $\sim$ 5--10 s.  Unlike the type II bursts from
the Rapid Burster, no relationship between the burst fluence and the
time to the next (or previous) burst has been reported\footnote{A
preliminary analysis of the BATSE catalog of over 3000 bursts from
\groj\ shows a statistically significant but weak correlation between
the burst fluence scaled to the persistent emission level and the time
to the next burst (Kommers et al. 1997).} (\cite{KouveliotouNature96};
\cite{StrickmanApJLett96}).

Using the first observations of \groj\ with the {\it Rossi X-ray
  Timing Explorer} (\xte), Swank (1996) noted that the bursts were
followed by a characteristic ``dip'' in the persistent emission level
which took a few minutes to recover.  No such dips were seen before
the bursts.  Subsequent \xte\ observations showed that the bursts
were sometimes preceded by steadily increasing variability in the
persistent emission, including ``micro'' and ``mini'' bursts
(\cite{GilesIauc6338}; \cite{LewinApJLett96}; \cite{GilesApJLett96}).

Extensive reviews of the Rapid Burster and its complex behavior can be
found in Lewin, van Paradijs, \& Taam (1993, 1995); see also
references therein.  Here we discuss the features of the type II
bursts from this source that are relevant to our comparison with
\groj.  The time intervals between type II bursts from the Rapid
Burster range from $\sim$ 10 s to $\sim$ 1 hr, with the shorter
intervals being more common.  Burst durations range from $\sim 2$ s to
$\sim 680$ s.  The burst repetition pattern is that of a relaxation
oscillator: the fluence of a type II burst is roughly proportional to
the time to the next burst (\cite{Lewin76}; Lewin et al. 1995).
Persistent X-ray emission between the type II bursts is observed
following long (duration $> 30$ s) bursts.  The persistent flux
emerges gradually after high-fluence bursts and decreases prior to the
next burst (\cite{marshall79}; \cite{vanparadijs79};
\cite{stella88a}).  These pre- and post-burst features are referred to
as ``dips'' in the persistent emission.  The spectrum of the
persistent emission is relatively soft during the dip just after a
burst.  It then rapidly increases in hardness, remaining hard for
$\sim$ 1--2 minutes before gradually decreasing to again become very
soft during the dip preceding the next burst (\cite{Stella}).  The
1--2 minute period of spectrally hard emission corresponds to a
``hump'' in the persistent emission light curve.

Lubin et al.  (1992) found ``naked eye'' quasi-periodic oscillations
(QPO) following 10 of 95 type II bursts from the Rapid Burster
observed with \exosat\ in August 1985.  The oscillations occurred {\it
  only} during the spectrally hard humps (which immediately
followed the post-burst dips).  The frequency of the oscillations
ranged from 0.039 to 0.056 Hz, with a period decrease of 30--50\%
observed over the $\sim 100$ s lifetime of the oscillations.  The
fractional root-mean-square (rms) variation in the oscillations was 5
to 15\%.  In 8 of the 10 cases, the ``naked eye'' oscillations were
accompanied by $\sim 4$ Hz QPO with rms variations of 6 to 19\%
(\cite{LubinNew}).

Several authors have already noted similarities between the bursts
from \groj\ and the Rapid Burster.  After the discovery of the bursts
from \groj, Kouveliotou et al. (1996) suggested that the release of
gravitational potential energy following an accretion instability
might be responsible for the bursts.  Lewin et al. (1996) made a
detailed comparison between the two systems.  Noting that both are transient
low-mass X-ray binaries in which the accretor is a neutron star, they
concluded that the bursts from \groj\ must be type II based on (i) the
hardness of the burst spectra, (ii) the lack of spectral evolution
during the bursts, (iii) the fact that the ratio of integrated energy
in the persistent emission to that in the bursts was initially too
small to allow for a thermonuclear burst mechanism, and (iv) the
presence of dips in the persistent emission following bursts
(\cite{LewinApJLett96}).  Subsequently, Sturner \& Dermer (1996)
reached the same conclusion.  The possibility of thermonuclear burning
in \groj\ has been discussed by Bildsten \& Brown (1997).

In this paper, we present another similarity between the Rapid Burster
and \groj.  Both sources show transient QPO during spectrally hard
emission intervals following bursts.

\section{Observations}
Since January 18, 1996, \xte\/\ has performed numerous observations of
\groj.  The data discussed here were taken by the PCA instrument and
are publicly available.  Between January 18 and April 26, 1996, there were
94 main bursts observed (\cite{GilesApJLett96}).  To refer to these
bursts individually, we number them sequentially from 1 to 94.

\begin{figure}
%\plotfiddle{f1.eps}{5.0in}{0}{100}{100}{-144}{0}
\plotone{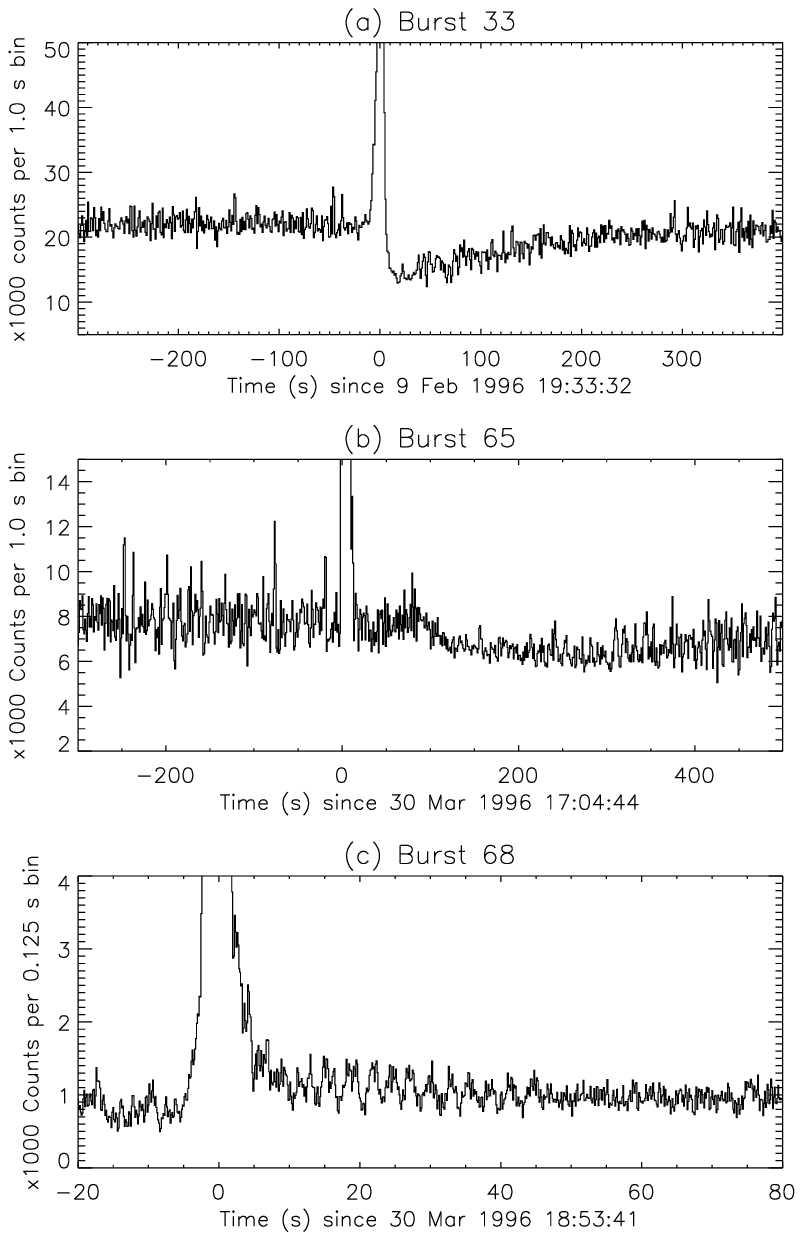}
\caption{Light curves of 3 bursts from \groj.  The horizontal and vertical scales
  are each different to highlight the feature of interest.  Panel (a)
  shows the ``dip'' following burst 33.  Panel (b) shows a
  ``shoulder'' after burst 65; it occurs before a slight ``dip''.  QPO
  with a mean frequency of $\sim$ 0.35 $\pm$ 0.02 Hz are present in
  the shoulder but are not apparent at this scale and time binning.
  Panel (c) shows an example of strong QPO with mean frequency 0.37
  $\pm$ 0.03 Hz occurring during the shoulder after burst 68.  The
  termination of the shoulder and the ``dip'' occur after the plotted
  time interval.
\label{FigLC}}
\end{figure}

The light curves of the bursts reveal a rich phenomenology in the
burst profiles.  Figure \ref{FigLC} (a) shows the ``dip'' in
persistent emission which follows some bursts, as first noted by Swank
(1996).  A broad ``shoulder'' (or ``plateau'') of emission above the
mean pre-burst level immediately follows some bursts, as shown in
Figure \ref{FigLC} (b).  The shoulder occurs {\it before} the
dip whenever both features are present.  Some bursts show
large-amplitude oscillations with frequencies $\sim 0.4$ Hz during
their shoulder, as shown in Figure \ref{FigLC} (c)
(\cite{KommersIauc6415}).

Oscillations can be seen during the shoulders of the light curves
immediately following at least 10 of the 94 bursts.  These ten bursts
were numbers 8, 12, 14, 18, 43, 53, 65, 68, 77, and 94.  Typically 5
to 15 cycles of the QPO are apparent, with as many as $\sim 25$ cycles
seen in the case of burst 65.  The low frequency of the oscillations
and the low numbers of cycles make it difficult to use Fourier power
spectra to study these QPO.  We have instead found the mean periods of
the oscillations by estimating the time intervals between successive
maxima in the count rates.

The mean frequency of the oscillations over the ensemble of bursts is
0.38 $\pm$ 0.04 Hz.  The mean frequency of the oscillations after
individual bursts varies from 0.35 Hz $\pm$ 0.02 Hz in burst 65 to
0.49 $\pm$ 0.03 Hz in burst 12.  (The uncertainty in these figures
represents the uncertainty in the mean frequency.)  The period of the
QPO following a given burst typically wanders non-monotonically about
a mean period by $\sim$ 20 percent over the lifetime of the
oscillations.  Using the $\chi^2$ statistic we exclude (at the
$2\sigma$ level or better in each of the 10 bursts) the null
hypothesis that the period between successive QPO pulses is constant.

To identify any overall tendency towards increasing or decreasing QPO
periods, we looked for bursts where the Spearman rank-order
correlation coefficient ($r_s$) indicated that the time intervals
between QPO pulses were correlated (or anti-correlated) with arrival
time.  In no case did we find a positive $r_s$, which would have
indicated an overall trend towards increasing periods.  For bursts 18,
53, 65 (see Figure 2), and 94, we found an anti-correlation at the
$2\sigma$ level or greater.  In these bursts the QPO period tends to
decrease by 10 to 20 percent over the lifetimes ($\sim$ 50 to
80 s) of the oscillations.

The fractional rms variations of the oscillations range from $\sim$ 5
to 13\%.  Using the counts to flux conversion $4.1 \times 10^{-12}$
erg cm$^{-2}$ s$^{-1}$ per count (total PCA band;
\cite{GilesApJLett96}) the rms variations in flux units range from
$1.2 \times 10^{-9}$ erg cm$^{-2}$ s$^{-1}$ to $1.4 \times 10^{-8}$
erg cm$^{-2}$ s$^{-1}$.

\begin{figure}
%\plotfiddle{f2.eps}{5.0in}{0}{100}{100}{-144}{0}
\plotone{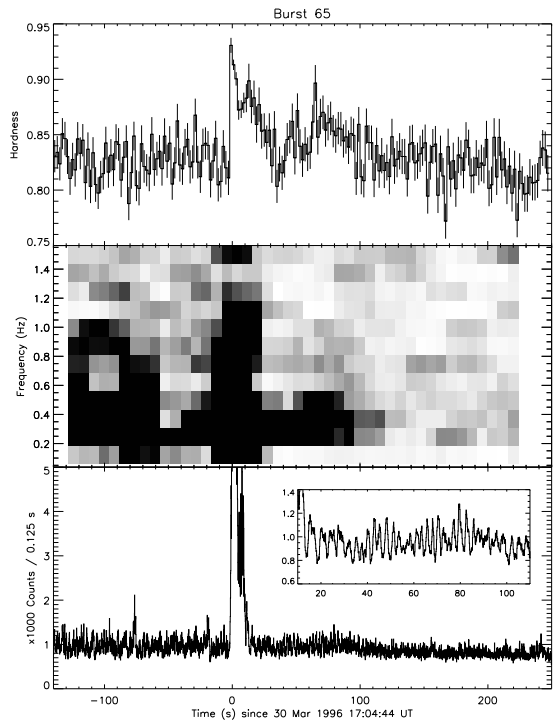}
\caption{Hardness, power spectrum, and intensity profile of burst 65.
  The top panel shows the ratio of the 3--5.6 keV count rate to the
  2--3 keV count rate.  (No instrumental dead time corrections have
  been applied, which may affect the hardness during the burst.  The
  hardness during the shoulder is not significantly affected by
  dead time.) The middle panel shows a power spectrum with higher
  powers represented by darker shades.  The lower panel shows the
  intensity profile.  Notice that the 0.35 $\pm 0.02$ Hz QPO occur
  after the burst, and the spectrum during their presence is
  relatively hard.  The inset shows the portion of the intensity
  profile during which the QPO were observed.  The data in the inset
  have been smoothed with a boxcar average to highlight the apparent
  beating between oscillations at $\sim$ 0.325 Hz and $\sim$ 0.375 Hz.
\label{FigHardness}}
\end{figure}

The shoulders during which the post-burst oscillations occur are
spectrally harder than the persistent emission.  Figure
\ref{FigHardness} (top panel) shows a hardness ratio for burst 65.
The hardness is defined as the count rate (averaged over 1.0 s) in the
3--5.6 keV range divided by that in the 2--3 keV range.  The middle
panel shows the power spectrum as a function of time.  Higher power
levels are shown with darker shades.  The presence of the QPO is shown
by a dark band at 0.4 Hz lasting from the end of the burst until $\sim
100$ s after the peak.  The bottom panel shows the burst intensity
profile.  The spectrum of the burst counts is clearly harder than that
of the persistent emission, but even after the main part of the burst
subsides the hardness ratios during the presence of the QPO (20 to 100
s after the burst) exceed those of the pre-burst persistent emission.

The QPO following burst 65 show a modulation envelope reminiscent of
``beating'' between oscillations at separate frequencies; see the
bottom panel of Figure \ref{FigHardness}.  The count rates shown in
the inset have been smoothed with a boxcar average to highlight the
envelope of the (roughly) 0.05 Hz ``beats''.  If this modulation
envelope can be attributed to simple beating, the two narrow-band
oscillations must occur concurrently and have roughly comparable
amplitudes.  An estimate of the two frequencies involved can be
obtained from the mean frequency $f_{ave} = (f_1 + f_2)/2$ and the
beat frequency $f_{beat} = |f_1 - f_2|$.  The mean frequency for the
$\sim 25$ cycles of QPO following this burst is 0.35 $\pm$ 0.02 Hz.
The two frequencies which are beating must then be roughly $f_1 =
f_{ave} - f_{beat}/2 = 0.325$ Hz and $f_2 = f_{ave} + f_{beat}/2 =
0.375$ Hz.  The average frequency wanders by about 20\%, so if beating
is responsible for the apparent modulation envelope then the
frequencies of the two narrow-band oscillations wander as well.  The
presence of a similar phenomenon following other bursts is difficult
to determine because fewer cycles are available.

Over the course of the January---May 1996 \xte\ observations, the
burst fluence, peak flux, and persistent emission level each decreased
(approximately linearly) by a factor of $\sim$ 4--5
(\cite{GilesApJLett96}).  The average frequency and rms amplitude of
the post-burst QPO show no significant ($> 2\sigma$) correlations with
any of these quantities.  QPO with nearly the same frequency are seen
in bursts for which the persistent emission level differs by a factor
of $\sim 4$.

Although our analysis focused on the December 1995--May 1996 outburst
of \groj, we note that post-burst oscillations were also detected
following a burst in June 1996.  This burst occurred when the source
temporarily resumed bursting activity at a lower level than the main
December 1995--May 1996 outburst (\cite{JahodaIauc6414}).  Kommers et
al.  (1996) reported 0.4 Hz oscillations during a $\sim 25$ s
``shoulder'' following the sixth of 7 bursts observed with the PCA on
June 4, 1996.  The fractional rms amplitude of these oscillations was
$25\pm 5$ percent.  A second large outburst from \groj\ has been in
progress since December 2, 1996, but we have not yet analyzed these
observations (\cite{KouveliotouIauc6530}; \cite{StarkIauc6548}).

\section{Discussion}
The $\sim$ 0.4 Hz oscillations following some bursts from \groj\ are
reminiscent of oscillations that have been seen following some type II
bursts from the Rapid Burster.  This likeness between the two sources
complements the comparisons made by Lewin et al. (1996).  Although it
is not certain that the post-burst oscillations are the same
phenomenon in both sources, the following similarities are worth
consideration.  (i) The post-burst oscillations follow bursts that do
{\it not} show the profiles or spectral evolution characteristic of
Type I (thermonuclear flash) bursts.  (ii) The oscillations occur
during a period of spectrally hard emission.  (iii) When an overall
change in QPO period is observed, it is a decrease: 10--20\% for
\groj, and 30--50\% for the Rapid Burster.  (iv) The fractional rms
amplitude of the oscillations is roughly 5--15\%.

There are some differences, however.  (i) The frequency of the
post-burst oscillations in \groj\ is roughly 0.4 Hz while in the Rapid
Burster, the post-burst oscillations have frequencies of 0.04 Hz and
are in some cases accompanied by 4 Hz QPO.  (ii) In \groj, the
oscillations occur {\it before} the dip in the persistent emission;
but in the Rapid Burster, they appear {\it after} the dip
(\cite{LubinNew}).

Cannizzo (1996a, 1996b) has shown that a simple numerical model of a
thermal-viscous accretion disk instability can reproduce some features
of the \groj\ bursts.  In his model, interplay between radial and
vertical energy transport in the disk causes oscillations in the ratio
of gas pressure to total pressure (denoted by $\beta$).  These
oscillations increase in amplitude and eventually become non-linear,
leading to a Lightman-Eardley instability (\cite{LightmanApJ74}).  For
specific choices of the inner and outer disk radii ($r_{inner} =
10^{7.5}$ cm, $r_{outer} = 10^{9}$ cm) and the form of the viscous
stress, integration of the time-dependent model reproduces the $\sim
1000$ s recurrence times and the $\sim 10$ s durations of the bursts
as well as the post-burst dips.  Cannizzo (1996b) also points out that
the increased variability {\it before} bursts may be related to the
oscillations in $\beta$ that precede the instability.  These
oscillations may occur in a variety of modes, but in his idealized
case appear to have frequencies of about 0.05 Hz
(\cite{CannizzoFluct96}).  It remains to be seen whether this model
can explain the oscillations reported here.

Abramowicz, Chen, \& Taam (1995) have shown that low-frequency
oscillations in mass flow can arise in accretion disk-corona systems.
In their model, the development of a Lightman-Eardley instability is
moderated by energy dissipation in the coronal region, which lies
above the thin disk.  They propose that the strong $\sim 0.042$ Hz
oscillations seen in the Rapid Burster may arise from such a
mechanism.  The frequency of these QPO may be relatively insensitive
to the accretion rate, since it is determined by several competing
factors including the radius of the inner accretion disk.  The
spectrum of the QPO is expected to be hard whenever the QPO originate
from the hot inner region (Abramowicz et al. 1995).

The common feature responsible for oscillatory behavior in these
models is the presence of a mechanism which couples the vertical
energy transport to the radial energy flux.  In Cannizzo's model, this
mechanism also acts to produce the bursts.  To make a quantitative
comparison of these models with the characteristics of the post-burst
oscillations in \groj\ and the Rapid Burster, one would have to know
what QPO frequencies are expected based on the physical parameters
that distinguish the two sources.  The magnetic field strength and
inner disk radius in particular should be quite different
(\cite{LewinApJLett96}; \cite{FingerNature96};
\cite{SturnerApJLett96}), which might account for the factor of $\sim
10$ difference in post-burst QPO frequency.  Detailed information of
this kind has not yet been presented for the models of Cannizzo
(1996a,b) or Abramowicz et al. (1995).

The presence of transient post-burst QPO is another similarity between
the bursts from \groj\ and the type II bursts from the Rapid Burster.
Combined with the comparison of Lewin et al. (1996), this similarity
is further evidence that the bursts from \groj\ are of type II.  If
the same disk instability is responsible for the type II bursts from
each of these sources, then the post-burst oscillations provide another
observational benchmark for models of the burst mechanism as well as
the QPO.

\acknowledgments J.\ M.\ K.\ and D.\ W.\ F.\ acknowledge support from
National Science Foundation Graduate Research Fellowships during the
preliminary phase of this research. J.\ M.\ K.\ acknowledges
subsequent support from a NASA Graduate Student Researchers Program
Fellowship NGT8-52816.  R.\ R.\ was supported by the NASA Graduate
Student Researchers Program under grant NGT-51368.  W.\ H.\ G.\ L.\
acknowledges support from NASA under grant NAG5-2046. J.\ v.\ P.\
acknowledges support from NASA under grant NAG5-2755.  C.\ K.\
acknowledges support from NASA under grant NAG5-2560.  We thank Dr. Ed
Morgan for helpful discussions and assistance with the \xte\ data
archive at MIT.

% This gets ready for the bibliography
\clearpage

% Here is the bibliography

\clearpage

%\clearpage

\end{document}